# Blood Glucose Measurement Based on Infra-Red Spectroscopy

Afarin Aghassizadeh, Mohammad Reza Nematollahi, Iman Mirzaie, Mehdi Fardmanesh
School of Electrical Engineering
Sharif University of Technology
Tehran, Iran
a.aghassizadeh@gmail.com, mohammadRn94@gmail.com, imanmirzaie@gmail.com, fardmanesh@sharif.edu

*Abstract*—**An algorithm based on PLS regression has been developed and optimized for measuring blood glucose level using the infra-red transmission spectrum of blood samples. A set of blood samples were tagged with their glucose concentration using an accurate invasive glucometer and analyzed using a standard FTIR spectrometer. Using the developed algorithm, the results of the FTIR spectroscopy of the samples were analyzed to find the glucose concentration in the samples. The obtained glucose concentration by the algorithm were in good agreement with the results obtained by the standard glucometer, and the mean estimation error was 7 mg/dL. This error is in the range of available commercial invasive meters.**

## I. INTRODUCTION

Diabetes mellitus is a type of metabolic disorder interfering with glucose metabolism in cells. Diabetes is due to either the weakness of the pancreas in producing insulin or failure of the cell membrane to respond to insulin properly[1].

Diabetes can cause many complications if not treated, causing cardiovascular disease, stroke, chronic kidney disease, foot ulcers, damage to the eyes and death. This is while glycemic control results in a normal life for patients and has been shown that it can increases life expectancy in diabetic patients[2].

According to World Health Organization (WHO) in 2017, 422 million adults were diabetic, worldwide[3]. To have a good control over glucose and avoiding the consequences of the diabetes, the patients need to control their blood glucose level at least four times a day[4].

Test is currently performed by invasive methods. The patient should pierce the skin (typically, on the finger) to draw blood, then apply the blood to a chemically active 'test-strip'. Then the device measures an electrical characteristic, and uses this to determine the glucose level in the blood.

As these techniques involve pricking on the fingertip, they are very painful for the patient and can cause serious blood infections. Moreover, the seemingly low cost strips used in the meters are a significant cost over time for the patients who their diabetes requires monitoring the blood glucose level in frequent intervals. These and other issues limits the number of sampling during a day and affects the glycemic control efficiency. Therefore, an accurate, painless, and easy-to-use device will help the patient to have more frequent testing and leads to a better glucose control and decreasing the long-term health care costs.

## II. INFRA-RED SPECTROSCOPY

There are many different approaches toward noninvasive glucometry. One of them is light spectroscopy. Spectroscopy is the study of the interaction between matter and electromagnetic radiation in term of absorption or scattering. It can be used as a measurement method for evaluating different compounds in a chemical sample both in visible and infrared spectrum.

Infrared spectroscopy exploits the fact that molecules absorb the energy significantly at their resonant frequencies that are characteristic of their structure. These frequencies are affected by the shape of the molecular potential energy surfaces, the masses of the atoms and the associated vibronic type couplings [5].

We chose Infra-Red (IR) region of the spectrum for our analysis because of the dominant absorption of the melanin and hemoglobin in the visible part of the spectrum which has a strong interfere with low glucose absorption in visible frequencies. On the other hand, we are also limited by the high absorption of water in higher wavelengths, which significantly decreases the signal to noise ratio. Besides, main glucose vibronic type absorption bands are in IR frequencies [6], [7].

## III. METHODOLOGY

To obtain a calibration model for blood glucose level using the IR spectroscopy information of blood samples, we used a set of blood samples tagged with their exact glucose concentration which was measured by an accurate invasive glucometer. The standard glucometer that we used was Accu-Check Active which it's technical specifications can be found in [8]. Then we obtained their transmission spectrum with an accurate and wideband Fourier Transform Infrared Spectrometer (FTIR). For this work we collected 47 blood samples from 3 different subjects. All of them were adults and had no history of diabetes. Two of the subjects were men and one of them was female. The blood sampling was done in different times in the day and contained glucose concentration from 84 mg/dl to 174 mg/dl.

## IV. DATA PROCESSING

### A. Preprocessing

Data-gathering methods are often loosely controlled, and collected data set must pass through some steps in order to be ready for further data mining procedures.

- Signal smoothing: We used a 10 point moving average filter in order to remove high frequency noises and smoothening the collected spectrums.

- Normalization: Normalization is used to unifying the effect of the variation range of independent variables in model identification. Since the range of values of blood volume varies in our measurements, normalization is essential in order not to underestimate the value of a measured spectrum.

- Mean centering: Standardizing the collected data so that they are centered on zero. Since the obtained algorithms in this work assume that our data is centered at 0, mean centering is essential for our work to be valid.

### B. Partial least square algorithm

Partial Least Squares (PLS) regression is a statistical method that identifies a linear model by trying to find the multidimensional direction in the input space (in our case the blood sample spectrums) that is associated with the maximum multidimensional variance direction in the output space (in our case the glucose level of blood samples).

PLS regression is particularly suited when the matrix of estimator has more variables than observations, and when there is multi collinearity among input space values. A more detailed discussion on PLS algorithms is given in [9].

Model complexity affects the fitting accuracy; more components result in better accuracy in fitting the available data set but destroys the prediction capability of the model. On the other hand, using few component will causes inaccuracy in the model. So a compromise on fitting accuracy and prediction capability is necessary. The optimum number of components is where the prediction error and fitting error have their minimum difference.

### C. K-fold cross validation

Cross validation methods are essential to evaluate the predictive algorithms. Methods based on residual evaluations do not give an indication of how well the model would work when new data is presented to the model. One way to overcome this problem is to divide the data into two sets of training data and test data. Then we can train the model with the training data set and then evaluate the prediction capabilities with test data set. This is the basic idea of the whole class of model evaluation methods called cross validation[10].

In K-fold cross validation the data set is divided into K subsets, and the identification process is repeated K times. Each time, one of the K subsets is used as the test set and the other subsets are put together to form a training set. Then the average error across all k trials is computed. The advantage of this method is that every data point gets to be in a test set exactly once. The variance of the results decrease as K increase [11].

In this work we chose K to be 5, and run the algorithm with just one component. We evaluated the difference between fitting error and predication error for each run, then increased the number of components used in PLS algorithm till we arrived at optimum number of components. The optimal number of components minimizes the difference between calibration error and predication error and it was obtained to be 7 for this work.

## V. RESULTS

After finding the optimum number of the components for PLS algorithm, we divided the collected data set into 75% training dataset and 25% test dataset to tune the model. The mean error of estimation was 7 mg/dL, which is in the range of available commercial invasive meters error [12].

The results of the model in estimating unknown blood glucose concentration in test dataset are shown in table I. The average in estimation of glucose concentration error is 7 mg/dL and the maximum error is 23 mg/dL.

TABLE I. ESTIMATED AND REAL GLUCOSE CONCENTRATIONS

| *Real Glucose Concentration (mg/dl)* | *Estimated Glucose Concentration (mg/dl)* |
|---|---|
| 93 | 96 |
| 123 | 100 |
| 93 | 101 |
| 96 | 102 |
| 101 | 102 |
| 95 | 104 |
| 101 | 104 |
| 106 | 108 |
| 103 | 109 |
| 99 | 109 |
| 113 | 115 |

## VI. CONCLUSION

In this work we used an algorithm based on PLS regression to measure blood glucose levels using IR transmission spectrum of blood samples. We gathered a set of blood samples with their exact glucose concentration and obtained their transmission spectrum using a standard FTIR spectrometer. For data analysis we used PLS regression with K-fold cross validation to find the best number of components in regression, which was obtained to be 7. This algorithm leads to an error of 7 mg/dl which is in the range of available commercial invasive glucometers.

This results can shows that blood spectroscopy can bring us closer to making a noninvasive glucometer.

The proposed methodology for finding the blood glucose here can also be used for estimation of other biological components such as hormones, proteins, lipids, water contents, and etc. in blood samples. Estimation of these components are not only important and notable, but also can be used to derive a better estimation of the glucose concentration in the blood.